# Intercultural Communication Strategies of a Technology Brand: A Comparative Quantitative Analysis of Xiaomi's Digital Marketing in China and Russia

Artem Novobritskii

**Abstract:** In the 21st century, the era of globalization, consumers are dispersed across the globe, and brands compete for their attention and loyalty, largely within the digital realm. This reality elevates the importance of effective communication and the transmission of product value across diverse cultural contexts. This study presents a comparative quantitative analysis of the digital marketing strategies of Xiaomi, a leading Chinese technology brand, on two major social media platforms: Sina Weibo in China and VK (VKontakte) in Russia. The research investigates how Xiaomi adapts its communication to align with local cultural values, as defined by the theoretical frameworks of Hofstede and Hall. Through a frequency analysis of text-based posts and emoji usage, this paper demonstrates the significant differences in Xiaomi's communication strategies in these two markets. The findings reveal that in China, a market characterized by high masculinity and low uncertainty avoidance, Xiaomi's messaging focuses on innovation, authority, and aspiration. In contrast, in Russia, a market with high uncertainty avoidance and lower masculinity, the brand's communication is more pragmatic, emphasizing tangible product benefits and building emotional connections. This study contributes to the field of intercultural digital marketing by providing empirical evidence of how a global brand adapts its communication strategies to different cultural contexts. The findings offer valuable insights for multinational corporations seeking to develop effective global marketing strategies in an increasingly interconnected world.

# 1. Introduction

The 21st century has witnessed a profound shift in the global economic landscape, marked by the meteoric rise of Chinese technology brands. Companies like Huawei, Alibaba, and Xiaomi have transitioned from domestic players to formidable global competitors, challenging the long-held dominance of Western multinational corporations [1] (P110-115). This expansion has been fueled by a combination of factors, including rapid technological innovation, competitive pricing strategies, and a deep understanding of digital marketing. However, as these brands venture into new international markets, they face a critical challenge: how to effectively communicate their brand identity and product value to consumers from diverse cultural backgrounds [2] (P217). In an increasingly interconnected world, where social media platforms have become the primary channels for brand-consumer interaction, the ability to navigate the complexities of intercultural communication is no longer a niche skill but a fundamental prerequisite for global success.

The proliferation of digital technologies has fundamentally transformed the nature of marketing. The traditional model of mass-market advertising has given way to a more personalized and interactive approach, where brands can engage directly with consumers in real-time. Social media platforms, in particular, have emerged as powerful tools for building brand communities, fostering customer loyalty, and driving sales. However, the global reach of these platforms also presents a unique set of challenges. A marketing message that resonates with consumers in one country may be ineffective or even offensive in another. This is because communication is not a universal process; it is deeply embedded within cultural frameworks that shape everything from language and non-verbal cues to the very structure of logical arguments. As a result, global brands are increasingly adopting a "glocal" approach to marketing, which involves adapting their global brand identity to the specific cultural nuances of local markets [3] (P163-170).

This paper presents a case study of Xiaomi, a leading Chinese technology brand that has achieved remarkable success in both its domestic market and internationally. Founded in 2010, Xiaomi has rapidly grown to become one of the world's largest smartphone manufacturers, with a business model that relies heavily on online sales and digital community engagement [4] (P96-97). In the first quarter of 2025, Xiaomi «regaining the pole position in Mainland China's smartphone market for the first time in a decade. With a market share of 19%» [5]. In Russia, according to data from MTS, between January and October 2024, Xiaomi ranked first in the

Russian smartphone market, accounting for 21% of all unit sales [6]. This makes its social media presence a critical component of its overall marketing efforts. By analyzing how this single, coherent brand identity is articulated in different cultural settings, we can gain valuable insights into the practical application of intercultural communication theories in the digital age. This study focuses on a comparative analysis of Xiaomi's digital marketing in China and Russia. This pair was selected for several strategic reasons. Firstly, both nations represent significant and growing markets for consumer electronics. Secondly, while they share strong economic ties and are often grouped within the BRICS geopolitical group, China and Russia possess profoundly different cultural profiles, providing a rich context for a comparative study of intercultural communication strategies.

This research is guided by the primary question: How does Xiaomi's digital communication strategy on China's Weibo differ from its strategy on Russia's VK? To answer this question, the study pursues several secondary objectives:

•To determine how these strategic differences reflect the underlying cultural dimensions of China and Russia as defined by the theoretical frameworks of Hofstede and Hall.

•To analyze the observable effects of these differing strategies on user engagement metrics, such as 'likes' and the use of different words and emojis.

•To extract actionable insights and theoretical implications for the broader field of intercultural digital marketing.

This paper is structured as follows. Section 2 provides a comprehensive review of the relevant literature on intercultural communication, digital marketing, and the globalization of Chinese brands. Section 3 outlines the methodology used in this study, including the data collection and analysis techniques. Section 4 presents the findings of the comparative quantitative content analysis. Section 5 discusses the theoretical and practical implications of the findings; conclusion.

## 2. Literature Review

A comprehensive understanding of Xiaomi's intercultural communication strategies requires a multi-faceted theoretical lens. This review of literature synthesizes key frameworks from intercultural communication, examines their application within the dynamic field of digital marketing, contextualizes the global rise of Chinese technology brands, and finally, analyzes the specific social media ecosystems of China and Russia. By integrating these diverse streams of research, this section establishes the theoretical and empirical foundation upon which the present study is built.

The study of intercultural communication, which explores how individuals from different cultural backgrounds interact and communicate, provides the essential theoretical underpinning for this research. The field, pioneered by scholars such as Edward T. Hall, posits that communication is not a universal constant but a variable deeply embedded within cultural frameworks that dictate everything from non-verbal cues and politeness norms to the very structure of logical arguments [7] (P117-125) For global brands, a failure to appreciate these differences can lead to marketing messages that are at best ineffective, and at worst, culturally insensitive or offensive. Two of the most influential frameworks for understanding these cultural differences are Geert Hofstede's Cultural Dimensions Theory and Edward T. Hall's theory of high-context and low-context cultures.

## 2.1. Hofstede's Cultural Dimensions Theory

A foundational framework for quantifying and understanding cross-cultural differences is Geert Hofstede's Cultural Dimensions Theory [8] (P8-16). Initially developed through a large-scale study of IBM employees across more than 50 countries, this model proposes that national cultures can be compared along several key dimensions of values. While the model has its critics and has been updated over time, it remains the most widely used framework for cross-cultural analysis in business and marketing studies. The six dimensions are:

- Power Distance Index (PDI): This dimension expresses the degree to which the less powerful members of a society accept and expect that power is distributed unequally. High PDI cultures exhibit a greater acceptance of hierarchical order, while low PDI cultures favor more consultative and democratic relations.
- Individualism vs. Collectivism (IDV): This dimension contrasts societies where the ties between individuals are loose (Individualist) with those where people are integrated into strong, cohesive in-groups (Collectivist).
- Masculinity vs. Femininity (MAS): Masculine cultures value achievement, heroism, assertiveness, and material rewards for success. In contrast, Feminine cultures prefer cooperation, modesty, caring for the weak, and quality of life.
- Uncertainty Avoidance Index (UAI): This dimension measures a society's tolerance for ambiguity and uncertainty. Cultures with high UAI scores maintain rigid codes of belief and behavior and are intolerant of unorthodox behavior and ideas. Low UAI societies maintain a more relaxed attitude in which practice counts more than principles.

•Long-Term Orientation vs. Short-Term Orientation (LTO): This dimension describes how societies prioritize their past traditions while dealing with the challenges of the present and future. Long-term oriented societies encourage thrift and efforts in modern education as a way to prepare for the future. Short-term oriented societies prefer to maintain time-honored traditions and norms while viewing societal change with suspicion.

•Indulgence vs. Restraint (IVR): This dimension addresses the extent to which societies allow for the gratification of basic and natural human drives related to enjoying life and having fun. Indulgent societies allow relatively free gratification, while restrained societies suppress gratification and regulate it by means of strict social norms.

For the purpose of this study, the cultural scores of China and Russia provide a critical backdrop for interpreting Xiaomi's marketing strategies. As illustrated in Table 1, there are both significant similarities and stark differences between the two countries.

| Cultural Dimension | China Score | Russia Score | Key Insights for Marketing Communication |
|---|---|---|---|
| Power Distance (PDI) | 80 | 93 | Both are highly hierarchical. Russia even more so. Marketing should respect authority, use formal language, and potentially leverage authoritative figures or brand status. |
| Individualism (IDV) | 20 | 39 | Both are collectivist cultures. China is strongly so. Marketing should emphasize group benefits, community, harmony, and social proof. Russia is moderately collectivist. |
| Masculinity (MAS) | 66 | 36 | China is a masculine, achievement-oriented society. Marketing can focus on success, winning, and being the best. Russia is more feminine, valuing quality of life, modesty, and consensus. |
| Uncertainty Avoidance (UAI) | 30 | 95 | This is the most dramatic difference. Russia has extremely high UAI, needing clear rules, structure, and guarantees. Marketing must be explicit, detailed, and risk- |

| | | | reducing. China's low UAI allows for more ambiguity and novelty. |
|---|---|---|---|
| Long-Term Orientation (LTO) | 87 | 81 | Both cultures are highly long-term oriented, valuing persistence, thrift, and future rewards. Marketing can focus on long-term value and brand legacy. |
| Indulgence (IVR) | 24 | 20 | Both are highly restrained societies. Marketing should not be overly frivolous or emotional. A focus on practical benefits over pure pleasure is likely more effective. |

Table 1: Comparative Analysis of Hofstede's Cultural Dimensions for China and Russia. (Source: Geerthofstede.com, 2025) [9]

## 2.2. Hall's High-Context and Low-Context Cultural Framework

Complementing Hofstede's value-based dimensions, Edward T. Hall's theory of high-context and low-context cultures focuses on the communication process itself [10] (P105-117).

•High-Context Cultures: In these cultures, communication is indirect, implicit, and relies heavily on the shared context, non-verbal cues, and the relationship between speakers. The message is often layered, and what is not said can be as important as what is said. Harmony and face-saving are paramount. China is a classic example of a high-context culture.

•Low-Context Cultures: In these cultures, communication is direct, explicit, and verbal. The message is encoded in the words themselves, and there is less reliance on shared understanding. Clarity, assertiveness, and getting straight to the point are valued. Germany and the United States are typically cited as low-context cultures.

Interestingly, both China and Russia are classified as high-context cultures. This implies that for both markets, communication should be relational and pay attention to non-verbal and symbolic elements. However, research indicates a crucial nuance: Russian culture, while high-context, allows for a greater degree of directness and expressiveness, particularly in informal settings, than many other high-context cultures [11] (P4-7). This paradox suggests that while Russians value relationships and context, they also appreciate straightforwardness once a relationship is established, a trait that may be linked to their high Uncertainty Avoidance. This hybrid nature makes the Russia-China comparison particularly insightful.

## 2.3. Social Media Landscapes in China and Russia

A critical component of any intercultural digital marketing analysis is a thorough understanding of the specific platforms on which the communication takes place. The global social media landscape is far from uniform; it is a fragmented ecosystem of platforms, each with its own unique history, features, user base, and cultural norms.

Sina Weibo (微博), which translates to "micro-blog," is one of China's most influential social media platforms [12] (P2). While often compared to Twitter due to its microblogging format, this comparison only scratches the surface of its functionality and cultural significance. Weibo is a hybrid platform that combines the real-time news and information dissemination of Twitter with the rich media and social networking features of Facebook. It is a vibrant and often chaotic space where users share news, follow celebrities and influencers (known as Key Opinion Leaders, or KOLs), discuss trending topics, and engage with brands [13] (P5).

For marketers, Weibo offers a powerful platform for reaching a massive and highly engaged audience. The platform's open and public nature allows content to spread rapidly, and its "trending topics" feature can turn a brand's message into a national conversation overnight. Marketing on Weibo often involves a multi-pronged strategy that includes creating engaging content, running targeted advertising campaigns, and, crucially, collaborating with KOLs who can lend their credibility and reach to a brand's message. Furthermore, Weibo has integrated a range of e-commerce features, allowing users to purchase products directly within the platform, thus blurring the lines between social media and online retail.

VKontakte, more commonly known as VK, is the dominant social media platform in Russia and many other Russian-speaking countries. Launched in 2006, it bears a strong resemblance to Facebook in its design and functionality, offering features such as personal profiles, groups, private messaging, and a news feed. However, VK has also developed its own unique identity, with a particularly strong emphasis on music and video sharing [14] (P44).

From a marketing perspective, VK provides a comprehensive set of tools for brands to engage with Russian consumers. Brands can create official pages, share content, run targeted advertising campaigns, and build communities around their products and services. The platform's powerful search and filtering capabilities allow for a high degree of audience segmentation. However, effective marketing on VK requires an understanding of the platform's specific cultural norms. Russian users tend to be more direct and expressive in their online interactions than users in many other high-context cultures. They also place a high value on authenticity and are quick to criticize brands that appear to be insincere or overly corporate. Therefore, a successful VK strategy often involves a more personal and community-oriented approach, with a focus on building genuine relationships with users.

# 3. Methodology

To investigate how Xiaomi adapts its digital communication strategy across different cultural contexts, this study employs a comparative quantitative content analysis. This methodological approach is designed to systematically and objectively identify specified characteristics of messages, allowing for a direct and empirical comparison of communication strategies across the two distinct platforms and cultural settings of China and Russia.

## 3.1. Data Collection

The data for this study were collected from Xiaomi's official, verified brand accounts on Sina Weibo ([https://weibo.com/xiaomishouji](https://weibo.com/xiaomishouji), «小米手机») and VK ([https://vk.ru/xiaomi](https://vk.ru/xiaomi), «Xiaomi Россия»). These accounts represent the brand's official voice and serve as the primary channels for its digital marketing communications in their respective countries. The sample includes all text-based posts published on these accounts up to September 30, 2025. This time frame was selected to provide a recent and relevant snapshot of Xiaomi's communication strategy, encompassing the promotion of new product launches and ongoing marketing campaigns. In total, 164 posts were collected - 100 from the Weibo account and 64 from VK - reflecting the differing posting frequencies on the two platforms. The text of each post was manually extracted and compiled into a dataset. For every post, key information such as the publication date, full text, and number of likes was systematically recorded.

## 3.2. Data Analysis Techniques

The data analysis was conducted using quantitative methods and proceeded through several clearly defined stages to ensure methodological rigor and analytical depth.

First, all collected posts from Xiaomi's Weibo and VK accounts were compiled into a unified dataset. Each entry included the publication date, full text and number of likes. This dataset served as the empirical basis for subsequent statistical processing and text analysis.

Second, the textual component of the data underwent preprocessing. This stage included lemmatization, which reduced words to their base or dictionary forms in order to treat morphological variants (e.g., *run*, *runs*, *running*) as a single lexical unit. This normalization step was crucial for increasing the accuracy and interpretability of frequency-based measures. Following lemmatization, a stop-word removal procedure was applied. Common functional words and non-content-bearing tokens (such as *the*, *a*, *is*) were excluded from the corpus to focus the analysis on semantically meaningful vocabulary directly relevant to Xiaomi's communication strategy.

Third, a frequency analysis was performed on the processed text to identify the most frequently occurring words and emojis in each dataset. This step provided a quantitative overview of the key linguistic and symbolic elements that define the brand's digital messaging in the Chinese and Russian markets. The comparison of these frequency distributions enabled the identification of cross-cultural differences in thematic emphasis and stylistic preferences.

Finally, descriptive statistical analysis was applied to summarize the structural features of the posts, including average word count and engagement metrics (likes per post). These statistics provided a contextual framework for interpreting the quantitative content patterns observed across platforms. Together, these analytical procedures yielded a systematic and data-driven understanding of Xiaomi's intercultural communication strategies, grounded in empirical linguistic and behavioral evidence.

### 3.3. Limitations of the Study

It is important to acknowledge the limitations of this research. Firstly, the sample, while sufficient for identifying broad patterns, is limited in time and may not capture seasonal variations in marketing campaigns. Secondly, the exclusion of visual content (images and videos) means that a significant dimension of social media communication is not analyzed. Visuals play a crucial role in brand storytelling and emotional appeal, and their exclusion is a notable limitation. Finally, the focus on a single brand, Xiaomi, means that the findings may not be generalizable to all technology brands or industries. Despite these limitations, the study's focused, comparative, and mixed-methods approach provides a robust and valuable snapshot of intercultural strategic adaptation in practice, offering a solid foundation for future research in this area.

## 4. Findings and Analysis

A preliminary analysis of the collected data reveals significant differences in posting frequency, post length, and user engagement between the two platforms. These differences, summarized in Table 2, provide the initial context for understanding the divergent strategies.

| Metric | China (Weibo) | Russia (VK) |
|---|---|---|
| Follower Count | 28 million | 1,3 million |
| Average Posts per Day | ~5.5 | ~0.7 |
| Average Words per Post | 60 | 55 |
| Average Likes per Post | 597 | 110 |
| Likes-to-Followers Ratio | 0.00002 | 0.00009 |

**Table 2: Comparative Overview of Xiaomi's Weibo and VK Accounts.**

The most immediate observation is the vast difference in audience size, with the Weibo account having more than 20 times the number of followers as the VK account. This is unsurprising, given that China is Xiaomi's home market. The posting frequency is also higher on Weibo, suggesting a more active and content-intensive strategy. Interestingly, while the average number of likes per post is significantly higher on Weibo in absolute terms, the likes-to-followers ratio is approximately four times higher on VK. This suggests that while the Russian audience is smaller, it is proportionally more engaged, at least in terms of this specific metric. This could be attributed to several factors, including the different algorithms of the platforms, the nature of the content, or a more dedicated and interactive community on the Russian platform. The average post length is roughly similar, indicating that the difference in strategy is not simply a matter of verbosity but lies in the content itself.

To understand the strategic focus of Xiaomi's messaging in each market, a frequency analysis of the most common words was conducted after lemmatization and the removal of stop words. The results, presented in Table 3, reveal two distinct thematic universes, each tailored to the cultural predispositions of its target audience.

| Rank | China (Top Words on Weibo) | Russia (Top Words on VK) |
|---|---|---|
| 1 | Xiaomi | Xiaomi |
| 2 | new | smartphone |
| 3 | series | redmi |
| 4 | screen | photo |
| 5 | LeiJunAnnualKeynote | shot |
| 6 | also | day |

| 7  | rear     | camera  |
| 8  | flagship | battery |
| 9  | phone    | pro     |
| 10 | super    | zoom    |

Table 3: Top 10 Most Frequent Words in Xiaomi's Weibo and VK Posts.

The vocabulary on Weibo is heavily skewed towards themes of innovation, newness, and technological superiority. The frequent use of words like "new," "series," "flagship," and "super" creates a narrative of constant progress and cutting-edge technology. This aligns perfectly with China's high Masculinity score (66), which indicates a culture that values achievement, success, and being the best. The marketing message is aspirational, positioning Xiaomi not just as a provider of functional devices, but as a leader at the forefront of technological innovation. The prominence of "screen" and "rear" points to a focus on specific, high-performance technical features.

Furthermore, the inclusion of "LeiJunAnnualKeynote" in the top words is highly significant. Lei Jun is the founder and CEO of Xiaomi, a charismatic and highly respected figure in the Chinese tech industry [15] (P57-59). The frequent reference to his keynote address leverages his personal authority and celebrity status, a strategy that is particularly effective in a high Power Distance culture like China (PDI score of 80). By associating the brand with a powerful and authoritative figure, Xiaomi reinforces its status and credibility. This strategy also caters to China's relatively low Uncertainty Avoidance score (30), as the focus on novelty and the "next big thing" appeals to a consumer base that is more open to ambiguity and new experiences.

In stark contrast, the vocabulary on VK is grounded in the tangible, the practical, and the user-centric. The most frequent terms are not about abstract concepts of innovation but about concrete user applications: "photo," "shot," "camera," "battery," and "zoom." The focus is on what the phone does for the user. This pragmatic approach directly addresses the most salient feature of the Russian cultural profile: its extremely high Uncertainty Avoidance Index (UAI) of 95. In a culture that is highly averse to ambiguity and risk, consumers are more persuaded by clear, factual information about a product's performance than by abstract promises of innovation. The emphasis on the camera's capabilities ("photo," "shot," "zoom") and the battery life ("battery," "day") provides tangible proof of the product's quality and reliability, thereby reducing the consumer's perceived risk.

This strategy is further reinforced by Russia's lower Masculinity score (36), which suggests a culture that places a higher value on quality of life, modesty, and consensus. The focus on practical benefits like taking good photos and having a long-lasting battery appeals to a desire for a more convenient and enjoyable user experience, rather than a need to have the newest and most powerful device. The strategy is not to sell an abstract vision of the future, but to provide a reliable tool for today. This focus on dependable performance is a classic strategy for high-UAI markets, where trust is built through evidence and predictability rather than aspirational hype.

Emojis, far from being mere decoration, have become a sophisticated form of non-verbal communication in the digital realm. They can convey tone, emotion, and meaning in a way that text alone often cannot. An analysis of the most frequently used emojis in Xiaomi's posts reveals another layer of cultural adaptation.

| Rank | China (Top Emojis on Weibo) | Russia (Top Emojis on VK) |
|---|---|---|
| 1 | ✅ | 🧡 |
| 2 | 👇 | 📷 |
| 3 | 👇 | ✨ |
| 4 | 🎉 | 🔷 |
| 5 | 📷 | 🔋 |

Table 4: Top 5 Most Frequent Emojis in Xiaomi's Weibo and VK Posts.

The dominant emojis on Weibo are functional and directive. The checkmark (✅) and pointing fingers (👇, 👇) act as navigational cues, guiding the user's attention in a fast-scrolling, information-dense feed. They are a form of visual shorthand, telling the user "look here," "this is important," or directing them to a link in the comments. This aligns with the platform's nature as a news and information hub, where users are scanning for information quickly. The party popper (🎉) signifies event-driven marketing, used to announce launches, sales, or the aforementioned keynote, reinforcing the sense of excitement and occasion. The camera emoji (📷) is present, but it is more of a generic signifier for "photo" rather than part of a broader emotional context. The overall emoji strategy on Weibo is one of efficiency and direction, designed to cut through the noise and guide user behavior.

The emoji usage on VK tells a different story, one focused on building emotional connection and reinforcing key messages. The most frequent emoji is the orange heart (🧡), a clear and direct attempt to build a sense of warmth, affection, and community with the audience. This is a relationship-building gesture, consistent with the more community-oriented nature of the VK platform. The sparkle emoji (✨) adds a touch of magic or special quality to the product, subtly enhancing its appeal. Crucially, the camera (📷) and battery (🔋) emojis serve as visual reinforcements of the key practical benefits highlighted in the text. They are not just generic symbols; they are direct visual echoes of the core value proposition being communicated - a great camera and a long-lasting battery. This use of emojis to visually confirm the product's tangible features is another tactic that serves to reduce uncertainty for the high-UAI Russian consumer, providing an extra layer of clear, unambiguous information.

## 5. Discussion and Conclusion

The findings of this comparative analysis provide a clear and compelling picture of how a global technology brand, Xiaomi, strategically adapts its digital communication to the distinct cultural landscapes of China and Russia. The quantitative and qualitative data converge to reveal two divergent narrative strategies, each meticulously tailored to the values, communication styles, and platform-specific norms of its target market. On China's Weibo, Xiaomi employs a strategy of aspirational innovation - authoritative in tone, focused on technological superiority, and reinforced by references to leadership and milestone events. In contrast, on Russia's VK, the brand adopts a more pragmatic and emotionally grounded approach, emphasizing tangible product benefits, community, and reliability. These distinct communication modes align closely with the respective cultural dimensions of each country: high Power Distance, Masculinity, and low Uncertainty Avoidance in China; and extremely high Uncertainty Avoidance and a more Feminine orientation in Russia.

The analysis of non-verbal elements further reinforces these cultural distinctions. Functional and directive emojis on Weibo serve as navigational tools, helping users process rapid streams of information, whereas the emotive and reinforcing emojis on VK cultivate relational warmth and trust. Together, these differences illustrate how Xiaomi effectively operationalizes cultural values into digital communication practices, using not only language but also visual and symbolic cues to connect with its audiences. Notably, despite the smaller audience on VK, the engagement intensity suggests that a more personalized, community-oriented strategy can foster deeper loyalty and sustained interaction.

The findings of this study have significant implications for intercultural communication theory, particularly in the digital age. They provide strong empirical support for the enduring relevance of Hofstede's and Hall's frameworks, demonstrating their applicability to modern social media environments. The alignment between Xiaomi's strategic choices and the cultural profiles of China and Russia shows that these classic models remain powerful tools for understanding and predicting communication behavior.

Practically, this research underscores the necessity of deep cultural localization in global marketing. A universal, one-size-fits-all approach is inadequate in the diversified environment of global social media. Brands must go beyond mere translation to craft messages that reflect the cultural values, communicative preferences, and behavioral motivations of their target audiences. Equally important is the recognition that platforms themselves embody unique social dynamics and technological affordances that shape communication. Effective global branding, therefore, requires both cultural intelligence and platform literacy. Xiaomi's case exemplifies how global consistency and local adaptability can coexist: by empowering local teams to tailor communication strategies to cultural expectations, the brand maintains a cohesive global identity while resonating authentically within each market.

While this study provides valuable insights, it also invites further exploration. Future research could adopt a longitudinal approach to track the evolution of Xiaomi's strategies over time or broaden the scope to include other brands and cultural contexts. Additionally, a deeper examination of visual and multimedia content would enhance understanding of how non-textual elements contribute to cross-cultural branding. Despite these limitations, the study offers both theoretical and practical contributions: it affirms that culture remains a fundamental determinant of communication behavior in the digital era and provides actionable guidance for global brands seeking to navigate the complexities of intercultural engagement online.

In essence, the case of Xiaomi demonstrates that successful global communication is not achieved through uniformity, but through cultural intelligence, adaptability, and respect for local sensibilities. A truly global brand is one that can speak in many voices while expressing one coherent identity - bridging cultures not by erasing their differences, but by skillfully weaving them into the fabric of its communication.